\documentstyle[aps,prb,epsf]{revtex}

\newcommand{\bra}[1]{\langle#1|}
\newcommand{\ket}[1]{|#1\rangle}
\newcommand{\e}{\begin{equation}}
\newcommand{\ee}{\end{equation}}

\begin{document}

\draft

%\wideabs{

\title{Perturbation of infinite networks of resistors}

\author{J{\'o}zsef Cserti${}^1$, Gyula D\'avid${}^2$ and Attila
Pir{\'o}th${}^3$}
\address{E{\"o}tv{\"o}s University, \\
${}^1$Department of Physics of Complex Systems,
${}^2$Department of Atomic Physics, \\
${}^3$Department for Theoretical Physics, \\
H-1117 Budapest, P\'azm\'any P{\'e}ter s{\'e}t\'any 1/A , Hungary}

\maketitle

%\newpage
\begin{abstract}

%abstract\\

The resistance between arbitrary nodes of infinite networks of
resistors is studied when the network is perturbed by
removing one bond from the perfect lattice.
A connection is made between the resistance and
the lattice Green's function of the perturbed network.
Solving Dyson's equation the Green's function and the resistance
of the perturbed lattice are expressed in terms of those of the
perfect lattice. Numerical results are presented for a square
lattice.
Our method of the lattice Green's function in
studying resistor networks can also be applied in the field of random
walks as well as electrical and mechanical breakdown phenomena 
in insulators, thin films and modern ceramics.

\end{abstract}

%\pacs{PACS numbers: }

%}

%\bigskip

%\pagestyle{plain}

%\newpage

\section{Introduction} \label{sec:intro}

The calculation of the resistance between two arbitrary nodes of
infinite networks of resistors is a well studied
subject\cite{vanderPol,Doyle,Venezian,Atkinson}.
Recently, it has been demonstrated  how the lattice Green's functions,
as an alternative method to using the principle of the superposition
of current distributions\cite{Venezian,Atkinson},
can be applied to this problem\cite{sajat-AmJP}.
The interested readers will also find several useful references there.

The analytical behavior of the lattice Green's function has been
extensively studied in condensed matter physics
over the past three decades\cite{Economou,Katsura};
impurities are often modeled by
a simple forms of perturbations (see Eq.\ (\ref{diad}) below).
Thus, the resistor network problem can be successfully tackled by
utilizing the strong connection between the two fields, in
particular through the use of the lattice Green's function\cite{sajat-AmJP}.
Our approach may also serve as a didactic introduction of
the Green's function method used in solid state physics as well
as field theory.
The perturbation series that show up in these fields are,
in general, infinite and need to undergo a tedious renormalization
procedure. In contrast, the perturbative solutions of
the Dyson equation can be summed up in closed form in our case,
due to the special form (\ref{diad}) of the perturbation.
Studying the analytical behavior of the closed-form Green's function
has also proven pedagogically useful in introductory courses
on solid state physics and field theory.

The conductivity of lattices with randomly
distributed defects has also been thoroughly investigated in the past 25
years within the framework of effective
medium theories\cite{Kirkpatrick,Koplik}
and by using the position-space renormalization-group method\cite{Redner}.
Below, we shall use the Green's function approach to
study a special case, i.e.\ to determine the resistance for a so-called
perturbed lattice that is
obtained by removing one bond from the perfect lattice
(corresponding to ``simple perturbation'' mentioned in
the previous paragraph).
Although this system is much less complex than those studied previously,
it has the great advantage of being analytically treatable and leading
to closed-form expressions for the resistances.

\begin{figure}
{\centerline{\leavevmode \epsfxsize=6.5cm \epsffile{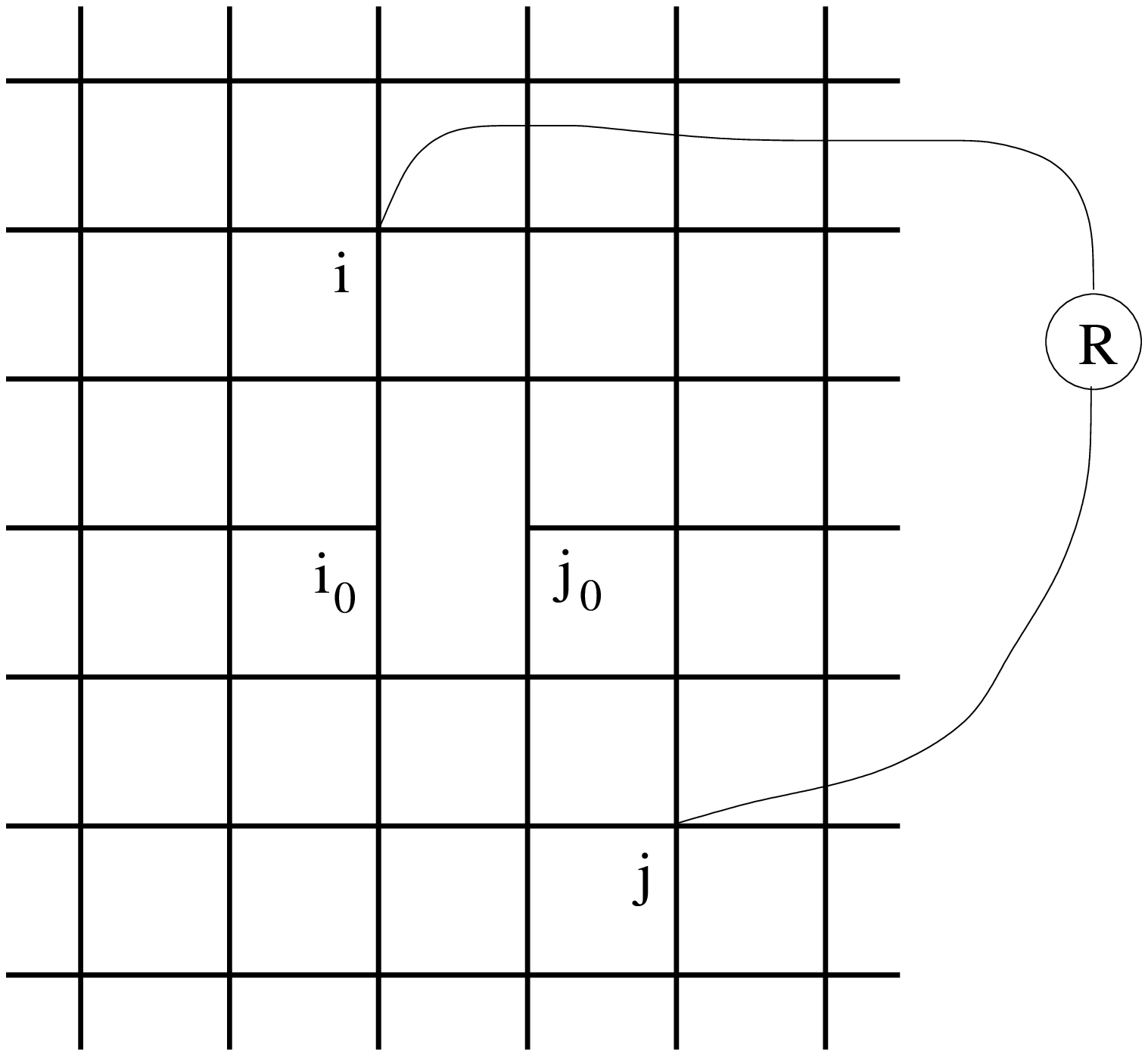 }}}
\caption{Perturbation of an infinite square lattice by removing one edge
between sites ${\bf r}_{i_0}$ and ${\bf r}_{j_0}$.
Each edge represents a resistance $R$.
The resistance $R(i,j)$ is measured between arbitrary lattice
points ${\bf r}_i$ and ${\bf r}_j$.
\label{abra-pert}}
\end{figure}

As an example (see  Fig.\ \ref{abra-pert}), consider an infinite
square lattice whose edges represent identical resistances $R$.
Removing one edge (bond) from this perfect lattice results
in a perturbed lattice.
Then, one can ask the resistance between two arbitrary nodes.
It is simple to find the resistance between the two ends of the missing
bond.
On one hand it is well known (see e.g.\ Aitchison's paper\cite{Aitchison})
that for a perfect square lattice the resistance between adjacent nodes is
$R/2$. On the other hand this resistance equals the parallel resultant of
$R$ and the resistance we wish to find.
Thus, it is obvious that the resistance between the two ends of the
missing bond in the perturbed lattice is $R$.

Finding the resistance between two arbitrary nodes in the
perturbed lattice is much more difficult.
In this paper we present a further application of the
Green's function method\cite{sajat-AmJP} to answer this question.
Our treatment of the problem is based on the idea used by
Kirkpatrick\cite{Kirkpatrick} for studying the transport in
inhomogeneous conductors. The Dirac notation of the vectors is used
throughout this paper.
Note that the derivation and our formulas presented below can be
straightforwardly expressed by the usual matrix-element formalism.
However, the Dirac notation seems to provide the most powerful and
elegant formulation of the problem. The readers who are unfamiliar
with this formalism will find sufficient background material in
several well-known textbooks, e.g.\ Schwabl\cite{Schwabl}.

According to Ohm's and Kirchhoff's laws, for a given current
distribution, the potentials on the lattice sites are given by a
Poisson-like equation involving a so-called lattice Laplacian
operator. This operator can be decomposed into two parts: one
associated with the perfect lattice and the other corresponding to
the perturbation. A relation between the resistance and the
Green's function for the perturbed lattice is derived.  The
problem thus reduces to the calculation of the Green's function
for the perturbed lattice which actually satisfies the Dyson
equation\cite{Economou}. As we shall see, the Dyson equation can
be solved exactly in the present problem. Then, an explicit
formula can be derived for the resistance in the perturbed lattice
in terms of the resistances between different nodes in the perfect
lattice. Our Green's function method proved to be a highly
effective technique and can successfully tackle the present
problem even in cases when the superposition
principle\cite{Venezian,Atkinson} faces hardly (or even
in-)surmountable difficulties.

\section{Green's function and resistance for perfect lattice}
\label{perfect-fej}

To establish the method and the notations that we shall use throughout
this paper, the case of a perfect lattice is reviewed first.
Consider a perfect $d$-dimensional infinite lattice made up of
identical resistances $R$.
All lattice points are specified by position vectors ${\bf r}$ given
in the form
\begin{equation}
{\bf r}=l_1{\bf a}_1 +l_2{\bf a}_2 + \cdots + l_d{\bf a}_d,
\label{d-lattice}
\end{equation}
where ${\bf a}_1,{\bf a}_2,\cdots,{\bf a}_d$ are independent primitive
translation vectors, and  $l_1, l_2,\cdots,l_d$ range through all
integer values.
We denote the potential at site ${\bf r}_i$ and the current
entering at lattice point
${\bf r}_i$ by $V({\bf r}_i)$ and $I({\bf r}_i)$, respectively.
According to Ohm's and Kirchhoff's laws,
we may write
\begin{equation}
\sum_{\bf n} \, \left[V({\bf r}_i)-V({\bf r}_i+{\bf n})\right]
=R \, I({\bf r}_i),
\label{aram-1}
\end{equation}
where the ${\bf n}$ are the vectors from site ${\bf r}_i$
to its nearest neighbors.
One can form two vectors, $V$ and $I$ from the potentials and the currents
at sites ${\bf r}_i$:
\begin{eqnarray}
V &=&  \sum_i \ket i V_i , \\
I &=&  \sum_i \ket i I_i ,
\end{eqnarray}
where  $V_i =V({\bf r}_i)$ and $I_i =I({\bf r}_i)$. Here
it is assumed that $\ket i$, associated with the site ${\bf r}_i$, forms
a complete orthonormal set, i.e.\ $\langle i | k
\rangle =
\delta_{ik}$ and $\sum_i \, | i \rangle \langle i | =1$.
Using the vectors $V$ and $I$, Eq.\ (\ref{aram-1}) can be rewritten as
\e
\sum_j \, \bigl(z \, \delta_{ij} -\Delta_{ij} \bigr)\, \bra j V
= R \, \bra i I,
\ee
where $z$ is the number of neighbors (coordination number) of each
lattice site (e.g.\ $z=2d$ for a $d$-dimensional hypercubic lattice),
and  $\Delta_{kl}$ is unity if the sites ${\bf r}_k$ and ${\bf r}_l$ are
nearest neighbors and zero otherwise.
The summation is taken over all lattice sites.
Premultiplying both sides of the above equation  by $\ket i$ and
summing over $i$, we have
\e
L_0 V = -R I,
\label{aram-2}
\ee
where  $L_0$ is the so-called
lattice Laplacian\cite{sajat-AmJP}:
\e
L_0=  \sum_{i,j}\,  \ket i \,
\bigl( \Delta_{ij} - z\, \delta_{ij} \bigr) \, \bra j.
\label{L0-def}
\ee

For further progress it is useful to define the lattice Green's function
of $L_0$. Similarly to the common definition used in the
literature\cite{Economou}, the lattice Green's function is defined by
\e
L_0 G_0 =-1.
\label{G0-def}
\ee
The properties of this lattice Green's function and its explicit form
in coordinate representation are given in Appendix \ref{app-G0}.
The solution of the Poisson-like Eq.\ (\ref{aram-2}) can then be given in
a simple form:
\e
V= -R L_0^{-1} I = R G_0 I.
\label{V-perfect}
\ee

To measure the resistance between two arbitrary sites
we assume that a current $I_0$ enters at site ${\bf r}_i$ and a current
$-I_0$ exits at site ${\bf r}_j$,
while the currents are zero at all other lattice points.
The current distribution may thus be written as
\e
I_m= I_0 \left(\delta_{mi} -\delta_{mj} \right)
\,\,\, \rm {for} \,\,\, \rm {all}\,\,\,\, m.
\ee
The potentials at different sites can be determined by inserting the
above-given current distribution into Eq.\ (\ref{V-perfect}), and
one finds
\e
V_k =\bra k V = R \bra k G_0 I = R \sum_m \, \bra k G_0 \ket m \, I_m =
R I_0 \bigl [G_0(k,i) -G_0(k,j) \bigr ],
\ee
where $G_0(n,m)= \bra n G_0 \ket m$ is the matrix element of the operator
$G$
in the basis $\ket n $.
The resistance between sites ${\bf r}_i$ and ${\bf r}_j$ is then
\e
R_0(i,j) = \frac{V_i -V_j}{I_0} = 2 R \bigl [G_0(i,i) -G_0(i,j) \bigr ],
\label{R0-def}
\ee
where we have made use of the symmetry properties of the Green's function
$G_0(i,j)$ given in Appendix \ref{app-G0}.
The same expression for the resistance in terms of the lattice Green's
function  was derived in Ref.\ \cite{sajat-AmJP}.

\section{Green's function and resistance for perturbed lattice.
Dyson's equation}
\label{perturbed-fej}

The current contribution $\delta I_i$ at site ${\bf r}_i$ due to the bond
$(i_0 \, j_0)$ is given by
\begin{eqnarray}
\delta I_i \, R &=& \delta_{i i_0}\left(V_{i_0} -V_{j_0} \right)
+ \delta_{i j_0}\left(V_{j_0} -V_{i_0} \right) \nonumber \\[1ex]
&=& \langle i | i_0 \rangle \bigl( \bra {i_0}-\bra {j_0}\bigr) V
+ \langle i | j_0 \rangle \bigl( \bra {j_0}-\bra {i_0}\bigr) V
\nonumber \\[1ex]
&=& \langle i | \bigl( \ket {i_0} - \ket {j_0} \bigr)
\bigl( \bra {i_0}-\bra {j_0}\bigr) V
=  \langle i | L_1 V, \nonumber
\end{eqnarray}
where  the operator $L_1$ is of a so-called
`dyadic' form:
\e
L_1 = \bigl(\ket {i_0}-\ket {j_0} \bigr)
\bigl( \bra {i_0}-\bra {j_0}\bigr).
\label{diad}
\ee
Here we used the fact that $\delta_{nm}=\langle n | m \rangle$.
Removing this bond from the perfect lattice the current $I_i$ at
site ${\bf r}_i$ is
given by
\e
{\left( -L_0 V\right)}_i -R\, \delta I_i = R\, I_i.
\ee
Thus, Ohm's and Kirchhoff's laws for the perturbed lattice can be written as
\begin{eqnarray}
L V &=& -R I, \,\,\,\, {\rm where}
\label{LV-1} \\[1ex]
L &=&  L_0 +L_1.
\label{L-def}
\end{eqnarray}
Equation (\ref{LV-1}) is similar to that found in the case of a
perfect lattice. However, the operator $L$ is now a sum of $L_0$
associated with the perfect lattice and a `perturbation' given by
$L_1$. Note that the same decomposition of the Laplacian operator
with the perturbation $L_1$ given by (\ref{diad}) was used by
Kirkpatrick.

Similarly to the perfect lattice, the Green's function $G$ for
the perturbed lattice is defined by
\e
L G =  -1.
\label{G-def}
\ee
The resistance for the perturbed lattice can be obtained in terms of the
Green's function much in the same way as for a perfect lattice.
Measuring the resistance between sites ${\bf r}_i$ and ${\bf r}_j$,
we assume that
a current $I_0$ enters at site ${\bf r}_i$ and a current $-I_0$ exits
at ${\bf r}_j$.
The currents are zero at all other lattice points.
Thus, the current distribution is
$I_m =  I_0 \left(\delta_{mi} - \delta_{mj} \right)$ for all $m$. 
From Eqs.\ (\ref{LV-1}) and  (\ref{G-def}) we have
$V_k = \bra k V =R \bra k G I = R \bra k G \sum_m \ket m \, I_m =
R \sum_m \bra k G \ket m \, I_m =R \sum_m G(k,m) \, I_m$. Inserting the
current distribution $I_m$ one finds
\e
V_k = R I_0 \bigl[ G(k,i) - G(k,j) \bigr].
\ee
Therefore, the resistance between sites ${\bf r}_i$ and ${\bf r}_j$
reads
\e
R(i,j) = \frac{V_i -V_j}{I_0} =
R \bigl[ G(i,i) -G(i,j) +G(j,j) - G(j,i) \bigr].
\label{R-def}
\ee
Note that unlike for a perfect lattice, $G(i,i) \ne G(j,j)$ since
translational symmetry is broken in the perturbed lattice. However, as we
shall see, $G(i,j) = G(j,i)$. The problem of finding the resistance
reduces to the calculation of the Green's function for the
perturbed lattice.

Perturbation theory for the Green's function is worked out in the
literature (see e.g.\ Economou's book\cite{Economou}).
Using  Eqs.\ (\ref{G0-def}), (\ref{L-def}) and (\ref{G-def})  yields
$\left(-G_0^{-1}+L_1\right)G =-1$. Premultiplying both sides of this
equation by $G_0$ one obtains the so-called Dyson's equation:
\e
G= G_0 + G_0 L_1 G .
\label{Dyson}
\ee
This is an equation for $G$ in terms of $G_0$ (which is assumed to be
known), and the perturbation $L_1$.  In general, the solution of the
Dyson equation can be found by iteration resulting in an infinite series:
\e
G=G_0 + G_0 L_1 G_0 +G_0 L_1 G_0  L_1 G_0 +\cdots .
\label{G-sor}
\ee
However, if $L_1$ has a special form -- such as that given
in Eq.\ (\ref{diad}) -- the Dyson equation can be solved exactly.
Now the perturbation $L_1$ is equal to the dyadic product of
the vector $\ket {i_0}-\ket {j_0}$ with itself.  In this case one may
apply the identity \e {\Bigl(\, A+ \ket x \bra y \, \Bigr)}^{-1} =
A^{-1} - \frac{A^{-1}\ket x \bra y A^{-1}}{1+\bra y A^{-1} \ket x}
\label{azonossag}
\ee valid for arbitrary vectors $\ket x $ and $\ket y $ whose
dimensions are the same as that of operator $A$, assuming that the
inverse of operator $A$ exists and the denominator $1+\bra y A^{-1}
\ket x\ne 0$.  This identity is readily proved by postmultiplying
Eq.\ (\ref{azonossag}) with $A+ \ket x \bra y $.  Using the above identity
with $A=L_0$, $\ket x = \ket {i_0}-\ket {j_0}$ and $\bra y = \bra
{i_0}-\bra {j_0}$ one obtains for the Green's function: \e G =
-{\left(L_0 + L_1 \right)}^{-1} = G_0 + \frac{G_0 \, \bigl(\ket
{i_0}-\ket {j_0} \bigr) \bigl( \bra {i_0}-\bra {j_0}\bigr)\, G_0}
{1-2\bigl[ G_0(i_0,i_0)-G_0(i_0,j_0) \bigr]}.
\label{G-op}
\ee
Below we shall see that the denominator in Eq.\ (\ref{G-op}) is
never equal to zero when $d>1$.
The matrix elements
of the operator $G$ can then be expressed with the matrix elements of
$G_0$ as \e G(i,j) = \bra i G \ket j = G_0(i,j) + \frac{\bigl[
G_0(i,i_0)-G_0(i,j_0) \bigr] \bigl[ G_0(i_0,j) -G_0(j_0,j)\bigr]}
{1-2\bigl[ G_0(i_0,i_0)-G_0(i_0,j_0) \bigr]}.
\label{G-matrix}
\ee

There is an alternative way to calculate $G$. This method is
analogous to that presented 
in Chapter 6.\ of Economou's book\cite{Economou} 
for the calculation of the Green's function. There the system was
condensed matter modeled by a periodic tight-binding Hamiltonian
with one impurity given by Dirac delta potential.

One can insert $L_1$ given by (\ref{diad}) into the series in
(\ref{G-sor}). Then the matrix elements of $G$ result in infinite
geometric series, which can be summed up exactly leading to the
same results as above.

Note that this geometric series as well as the perturbative
solution of the Dyson equation given by Eq.\ (\ref{G-sor}) is not
necessarily convergent, especially in the case of dyadic
perturbations with $|\bra y A^{-1} \ket x| \ge 1$. However, Eq.\
(\ref{azonossag}) is still valid even for $|\bra y A^{-1}\ket x| >
1$.

It is clear from $G_0(i,j) =G_0(j,i)$ that $G(i,j)$ is also a symmetric
matrix, i.e.\ $G(i,j) =G(j,i)$.
>From  Eqs.\ (\ref{R-def}) and (\ref{G-matrix}) the resistance
between ${\bf r}_i$ and ${\bf r}_j$ can be obtained in terms of the matrix
elements of $G_0$:
\begin{eqnarray}
\frac{R(i,j)}{R} &=& G(i,i) + G(j,j) -2G(i,j)  \nonumber\\[1ex]
&=&
2\left[G_0(i,i) - G_0(i,j)\right] +
\frac{
{\bigl[ G_0(i,i_0)-G_0(i,j_0)-G_0(j,i_0)+G_0(j,j_0) \bigr]}^2
}{1-2\left[G_0(i_0,i_0)-G_0(i_0,j_0)\right]}.
\end{eqnarray}
Using  Eq.\ (\ref{R0-def}) this formula can be rewritten in terms of
the resistances in a perfect lattice:
\e
R(i,j) = R_0(i,j) +
\frac{
{\bigl[ R_0(i,j_0)+R_0(j,i_0)-R_0(i,i_0)-R_0(j,j_0) \bigr]}^2
}{4\bigl[ R-R_0(i_0,j_0) \bigr]}.
\label{perturb-R}
\ee
This is our final result for the resistance between two arbitrary
nodes ${\bf r}_i$ and ${\bf r}_j$ of the perturbed lattice in which
the bond $(i_0 \, j_0)$ is removed.

It is easy to calculate the resistance between sites ${\bf r}_{i_0}$ and
${\bf r}_{j_0}$ for a $d$-dimensional hypercubic lattice.
For symmetry reasons\cite{sajat-AmJP} the
resistance between ${\bf r}_{i_0}$ and ${\bf r}_{j_0}$
in a perfect lattice is
$R_0(i_0,j_0)=R/d$ if $d>1$. Then, using Eq.\ (\ref{perturb-R}),
the resistance between the two ends of the removed bond is
$R(i_0,j_0)=R/(d-1)$.
For a square lattice ($d=2$) the resistance is $R$ as
mentioned in the introduction.
It also follows that the denominator in
Eq.\ (\ref{perturb-R}) is never equal to zero, and this is necessarily
true for each previous equation that contains this term.

Note that the explicit form of the lattice Laplacian $L_0$ defined
in Eq.\ (\ref{L0-def}) was not used in the derivation of
Eq.\ (\ref{perturb-R}). Thus, our final result for the resistance
in the perturbed lattice is valid for any lattice structure
in which each unit cell has only one lattice site.
This is the case, for example, for triangular lattices.

Finally, it is worth mentioning some possible applications of
our method and results. In Doyle and Snell's book\cite{Doyle} the
connection between random walks and electric networks is
presented. Thus, our result can also be extended to the random
walk problem in a perturbed lattice. Furthermore, the highly
effective method of the lattice Green's function in studying
resistor networks also provides a pedagogically useful way to
familiarize students with the notions of the Dyson equation and
the Green's function.

When more than one bond is missing from the perfect lattice, our
method (outlined above) can be iterated. It can be shown that our
final result (\ref{perturb-R}) is still valid provided the values
of $R_0$ of the perfect lattice are replaced by the values of the
resistances obtained from the previous step of the iteration.
Thus, lattices with more complex defects can be studied
analytically with our method. An example is the so-called
crack-type defects arising in several fields -- such as electrical
and mechanical breakdown phenomena in insulators, thin films and
modern ceramics\cite{Duxbury}. The current distribution around
cracks can be easily calculated from the Green's function. This
problem was studied numerically as well as analytically in the
continuum limit (for a recent paper see e.g.\ Boksiner and
Leath\cite{Boksiner} and references therein). Application of our
method to investigate this problem is in progress.

\section{Numerical results}
\label{num-res}

\begin{figure}
{\centerline{\leavevmode \epsfxsize=8.5cm \epsffile{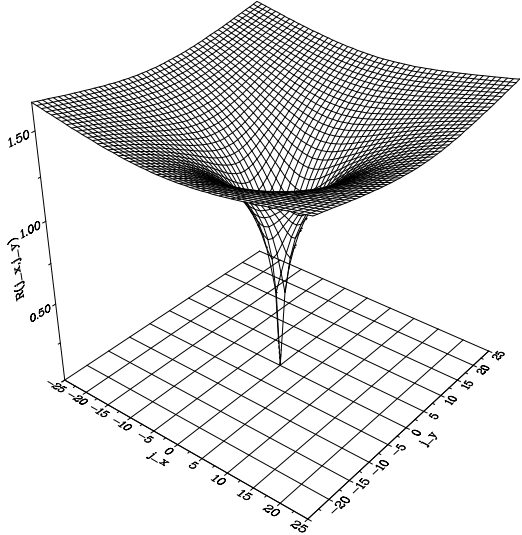}}}
\caption{The resistance $R_0(i,j)$ between nodes $i=(0,0)$ and
$j=(j_x,j_y)$ as a function of $j_x$ and $j_y$ for a perfect square lattice.
\label{abra-perfect}}
\end{figure}

Below we shall present some numerical results for square lattices.
Recurrence relations for the resistances  between arbitrary nodes
were derived for perfect square lattices in Ref.\ \cite{sajat-AmJP}.
These recurrence formulas provide a very simple and effective tool
for the calculation of the resistance. For convenience, we invoke them:
\begin{eqnarray}
R_0(m+1,m+1) &=& \frac{4m}{2m+1}\, R_0(m,m)-\frac{2m-1}{2m+1}\,
R_0(m-1,m-1),
\nonumber \\
R_0(m+1,m) &=& 2 R_0(m,m)-R_0(m,m-1),
\nonumber \\
R_0(m+1,0) &=& 4 R_0(m,0)-R_0(m-1,0)-2R_0(m,1),
\nonumber \\
R_0(m+1,p) &=& 4 R_0(m,p)-R_0(m-1,p)-R_0(m,p+1)- R_0(m,p-1)
\,\,\,\, {\rm if} \,\,\,\, 0< p < m,
\label{recur-sq}
\end{eqnarray}
where the indices $m$ and $n$ of $R_0(m,n)$ (not to be confused with
$i$ and $j$ above) stand for the {\it relative coordinates}
of the two sites on the square lattice.
Since the exact values of $R_0(1,0)=R/2$ and
$R_0(1,1)=2R/\pi$ are known\cite{sajat-AmJP}
(obviously $R_0(0,0)=0$), one can calculate the resistance exactly
for arbitrary nodes by using the above recurrence formulas.
It is interesting to note that using these formulas one has, in units
of $R$
$$R_0(50,50) = \frac
{6400711399252571342562758751832284129928}
{1089380862964257455695840764614254743075}\,
\frac{1}{\pi}\approx 1.87025,$$
which can be found quickly (within a few second) using the program MAPLE.
In Fig.\ \ref{abra-perfect} the resistance $R_0(i,j)$
between the node $i=(0,0)$ and $j=(j_x,j_y)$ is plotted
as a function of $j_x$ and $j_y$ for a perfect square lattice.
One can see from the figure that by increasing the distance between
the nodes $i$ and $j$ the rotational symmetry of the
resistance is more and more recovered.
Indeed, in Ref.\ \cite{sajat-AmJP} it was shown that
for large values of $j_x$ or/and $j_y$, the asymptotic form of
the resistance is
\begin{equation}
R(j_x,j_y)= \frac{R}{\pi} \left( \ln \sqrt{j_x^2+j_y^2} +\gamma + \frac{\ln
8}{2} \right),
\label{R-d2-asym}
\end{equation}
where $\gamma = 0.5772\dots$ is the Euler-Mascheroni
constant.

\begin{figure}
{\centerline{\leavevmode \epsfxsize=8.5cm \epsffile{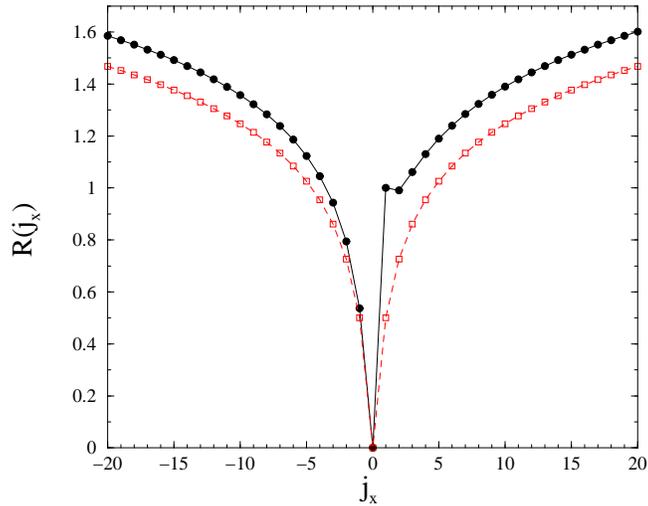 }}}
\caption{In an infinite square lattice the origin of the coordinate
system is at site ${\bf r}_i$, i.e.\ ${\bf r}_i=(0,0)$.
The ends of the removed bond are at ${\bf r}_{i_0}=(0,0)$ and
${\bf r}_{j_0}=(1,0)$.
The resistances on the perturbed and the perfect square lattices are
measured between sites ${\bf r}_i$ and ${\bf r}_j$, where
${\bf r}_j=(j_x,0)$. They are
plotted as functions of $j_x$ (filled circles and empty squares,
respectively).
The solid and dashed lines are just for guiding the eyes.
\label{abra-R-0}}
\end{figure}

On the perturbed square lattice the resistance can be calculated from
Eq.\ (\ref{perturb-R}).
As an example we show results when site ${\bf r}_i$ is fixed and
${\bf r}_j$ is moved along the line of the removed bond.
The resistance is measured between nodes ${\bf r}_i$ and ${\bf r}_j$,
where ${\bf r}_i=(0,0)$ and ${\bf r}_j=(j_x,0)$.
The two ends of the removed bond are ${\bf r}_{i_0}=(0,0)$ and
${\bf r}_{j_0}=(1,0)$. In Fig.\ \ref{abra-R-0} the resistances for the
perfect and the perturbed lattices are plotted as functions of $j_x$.
One can see that the resistance is always larger in the perturbed
lattice than in the perfect lattice. This intuitively obvious fact
follows from the positivity of the second term
in Eq.\ (\ref{perturb-R}). It can also be
seen from the figure that increasing the distance between nodes
${\bf r}_i$ and ${\bf r}_j$ the resistance tends to that of the perfect
lattice. The resistance is not symmetric
under the transformation $j_x \rightarrow -j_x$  because
translational symmetry is broken in the perturbed lattice.
\begin{figure}
{\centerline{\leavevmode \epsfxsize=8.5cm \epsffile{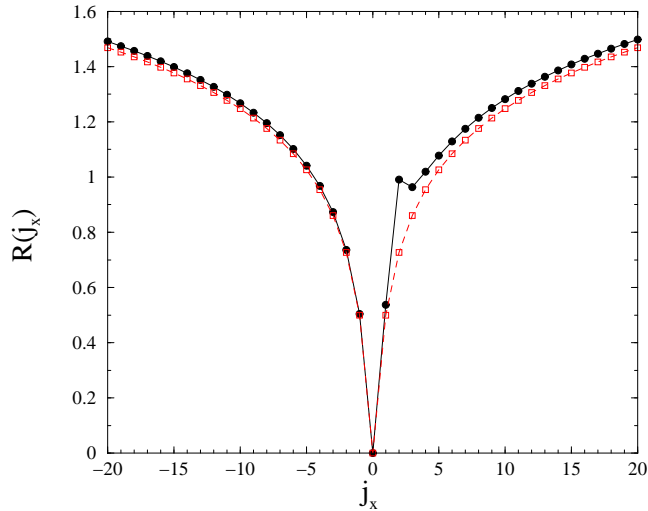 }}}
\caption{The same as Fig.\ \ref{abra-R-0} except that the ends of the
removed bond are at  ${\bf r}_{i_0}=(1,0)$ and ${\bf r}_{j_0}=(2,0)$.
\label{abra-R-1}}
\end{figure}
The configuration is the same in Fig.\ \ref{abra-R-1}  as
in Fig.\ \ref{abra-R-0} except that ${\bf r}_{i_0}={\bf r}_i +(1,0).$
Similar features are observed in the two cases.
However, the resistance on the perturbed lattice
tends to that of the perfect lattice more rapidly
in Fig.\ \ref{abra-R-1} than in Fig.\ \ref{abra-R-0} as $j_x$ increases.

Finally, it worth mentioning that a similar analysis of the resistance can
be carried out for simple cubic lattices.
Recently, Glasser et al.\ \cite{Glasser-3} have expressed the exact values
for the cubic lattice Green's functions in terms of gamma functions.
Using their results one can calculate the resistance for perfect simple
cubic lattices and for perturbed lattices.

\acknowledgements

The authors wish to thank S.\ Redner, L. Glasser and P.\
Gn{\"a}dig
for helpful discussions.
This work was supported by the EU.\ RTN within the programme
``Nanoscale Dynamics, Coherence and Computation'',
and the Hungarian  Science Foundation OTKA  TO25866 and TO34832.

\appendix
\section{Explicit form of the lattice Green's functions for
perfect lattice structures}
\label{app-G0}

In this section we shall present the symmetry properties of the lattice
Green's function $G_0$, and derive its explicit form in coordinate
representation for perfect lattice structures.

From Eq.\ (\ref{L0-def}) we have
$L_0(i,j)=\bra i L_0 \ket j = L_0(j,i)$. Therefore,
\e
G_0(i,j)=\bra i G_0 \ket j = G_0(j,i).
\ee
Thus,  $G_0(i,j)$ is a symmetric matrix.
The perfect lattice is translationally invariant, so  $G_0(i,j)$
depends only on the relative coordinates of sites ${\bf r}_i$
and ${\bf r}_j$.
Thus we have
\e
G_0(i,i)=G_0(j,j).
\label{azonos-hely}
\ee

Let us turn to the derivation of the coordinate representation of $G_0$.
Consider a finite part of the lattice constructed by repeating the unit
cell (say $N_i$ times) along the
unit vectors ${\bf a}_i$, where $i=1,2,\cdots,d$.
Define the Fourier transform of the basis $\ket i$ by
\e
|{\bf k}) = \frac{1}{\sqrt{N}} \sum_i e^{i{\bf k}{\bf r}_i}\, \ket i .
\ee
The lattice sites ${\bf r}_i$ are specified by Eq. (\ref{d-lattice}),
and the vectors ${\bf k}$ belonging to the first  Brillouin zone (BZ)
are given by
\begin{equation}
{\bf k} = \frac{m_1}{N_1}{\bf b}_1 + \frac{m_2}{N_2}{\bf b}_2
+ \cdots + \frac{m_d}{N_d} {\bf b}_d,
\label{k-vector}
\end{equation}
where $m_1,m_2,\cdots\,m_d$ are integers such that
$-N_i/2 \leq m_i \leq  N_i/2$
for $i=1,2,\cdots,d$, and the ${\bf b}_j$ are the reciprocal lattice vectors
defined by ${\bf a}_i {\bf b}_j = 2\pi \delta_{ij}$, $i,j=1,2,\cdots,d$.
The total number of lattice sites is $N=N_1 N_2 \cdots N_d$.
Periodic boundary conditions are assumed at the boundary of the finite
lattice, and in the final results the limit $N \rightarrow \infty$ is taken.
It is also assumed that $N_i$ is an even integer, which will be irrelevant
in
the limit $N \rightarrow \infty$.
The mathematical description of crystal lattices and the concept of
the Brillouin zone can be found in many books on
solid state physics\cite{szilfiz}.
To distinguish the basis vectors $\ket i$ from their Fourier transforms
$|{\bf k})$, parentheses will be used instead of angled brackets.
Using the so-called lattice summation rules\cite{szilfiz}
\begin{eqnarray}
\sum_i \, e^{i({\bf k}-{\bf k^\prime}){\bf r}_i} &=&
N\, \delta_{{\bf k}{\bf k}^\prime},
\\
\sum_{\bf k} \, e^{i{\bf k}({\bf r}_i - {\bf r}_j)} &=& N\, \delta_{ij},
\end{eqnarray}
it can be shown that the basis $|{\bf k})$ is a complete
orthonormal set, i.e.\ $({\bf k}|{\bf k}^\prime ) =
\delta_{{\bf k}{\bf k}^\prime}$ and
$\sum_{\bf k} |{\bf k})({\bf k}| =1$ provided the basis $\ket i$ is
complete and orthonormal. It is also clear that
\e
\bra i {\bf k}) = \frac{1}{\sqrt{N}} \,e^{i{\bf k}{\bf r}_i}.
\ee
The Laplacian operator $L_0$ for a perfect
lattice is a diagonal matrix in the basis $|{\bf k})$ since
\begin{eqnarray}
({\bf k}|L_0|{\bf k}^\prime ) &=& \sum_{ij} ({\bf k} \ket i \,
\bigl(\Delta_{ij} - z \, \delta_{ij}\bigr)\, \bra j {\bf k}^\prime )
= \frac{1}{N}\,  \sum_{ij} e^{-i{\bf k}{\bf r}_i} \,
\bigl(\Delta_{ij} - z \, \delta_{ij}\bigr)\,
e^{-i{\bf k^\prime}{\bf r}_j} =
- \varepsilon({\bf k}) \, \delta_{{\bf k}{\bf k}^\prime},
\end{eqnarray}
where
\e
\varepsilon({\bf k}) = z - \sum_{\bf n} e^{i{\bf k}{\bf n}},
\ee
and  vectors ${\bf n}$ point from some lattice point into its
the nearest neighbors.

In coordinate representation the matrix element of the Green's function
$G_0$ can be written as
\e
G_0(i,j) = \bra i G_0 \ket j =
\sum_{{\bf k}{\bf k}^\prime}\, \bra i {\bf k})
({\bf k} | G_0 | {\bf k}^\prime ) ({\bf k}^\prime \ket i =
\frac{1}{N}\, \sum_{{\bf k}{\bf k}^\prime}\,
({\bf k} | G_0 | {\bf k}^\prime )
e^{i({\bf k}{\bf r}_i-{\bf k^\prime}{\bf r}_j )}.
\ee
In the third step the completeness of the basis $|{\bf k})$ has been used.
Since $G_0 = -L_0$, one finds
\e
G_0(i,j) =\frac{1}{N}\, \sum_{{\bf k} \in {\rm BZ}} \,
\frac{ e^{i{\bf k}({\bf r}_i-{\bf r}_j )}}
{\varepsilon({\bf k})}.
\ee
Note that the toroidal and cylindrical geometry cases
discussed by Jeng\cite{Jeng} can be straightforwardly treated
by direct summation over ${\bf k}$.
For infinite lattices (in the limit $N \rightarrow \infty$)
the discrete ${\bf k}$
summation can be substituted by an integral
\cite{szilfiz}:
\e
\sum_{{\bf k} \in {\rm BZ}}  \rightarrow
\frac{N v_0 }{{(2\pi)}^d}\int_{{\bf k} \in {\rm BZ}} \, d^d {\bf k},
\ee
where $v_0$ is the volume of the unit cell of the lattice.
Thus, the lattice Green's function in coordinate representation becomes
\cite{sajat-AmJP}
\begin{equation}
G_0(i,j) = v_0\int_{{\bf k} \in {\rm BZ}} \,
\frac{d^d {\bf k}}{{(2\pi)}^d}\,
\frac{e^{i{\bf k}({\bf r}_i -{\bf r}_j)}}{\varepsilon({\bf k})}.
\label{G-veg}
\end{equation}
One can see that $G_0(i,j)$ depends only on the relative coordinates
of sites ${\bf r}_i$ and ${\bf r}_j$, as stated in Eq.\
(\ref{azonos-hely}).
For example,  $z=2d$ for the $d$-dimensional hypercubic lattice, and
one finds
$\varepsilon({\bf k}) = 2\left(d- \sum_{i=1}^d \cos {\bf k} {\bf
a}_i\right)$.
For body centered cubic and face centered
cubic lattice structures  see e.g.\ the introduction of the lattice
Green's function by Katsura et al.\ \cite{Katsura}, and for
triangular and hexagonal lattices Ref.\ \cite{sajat-AmJP}.

%\newpage

%\bibliography{cikkek}
%\bibliographystyle{/usr/lib/texmf/texmf/tex/revtex/osa}

\end{document}